\documentclass{taira}

\usepackage{xcolor}
\usepackage{graphicx}
\usepackage{amsmath,comment,xcolor}
\usepackage{longtable,tabularx}
\newcommand{\sqdiamond}[1][fill=black]{\tikz [x=1.2ex,y=1.85ex,line width=.1ex,line join=round, yshift=-0.285ex] \draw  [#1]  (0,.5) -- (.5,1) -- (1,.5) -- (.5,0) -- (0,.5) -- cycle;}%
\newcommand{\MyDiamond}[1][fill=black]{\mathop{\raisebox{-0.275ex}{$\sqdiamond[#1]$}}}

\usepackage{tikz}
\usetikzlibrary{shapes}
\newcommand{\markerone}{\raisebox{0.0pt}{\tikz{\node[scale=0.6,regular polygon, circle, draw = {rgb,255:red,0; green,114; blue,189},line width=0.3mm, fill={rgb,255:red,0; green,114; blue,189}](){};}}}
\newcommand{\markertwo}{\raisebox{0.0pt}{\tikz{\node[scale=0.45, regular polygon, regular polygon sides=3, fill={rgb,255:red,126; green,47; blue,142}](){};}}}
\newcommand{\markerthree}{\raisebox{0.0pt}{\tikz{\node[scale=0.45,regular polygon, regular polygon sides=3,fill={rgb,255:red,119; green,172; blue,48},rotate=-90](){};}}}
\newcommand{\markerfour}{\raisebox{0.0pt}{\tikz{\node[scale=0.45,regular polygon, regular polygon sides=3,fill={rgb,255:red,77; green,190; blue,238},rotate=-180](){};}}}
\newcommand{\markerfive}{\raisebox{0.0pt}{\tikz{\node[scale=0.55,regular polygon, regular polygon sides=4, draw = {rgb,255:red,237; green,177; blue,32}, line width=0.3mm, fill={rgb,255:red,255; green,255; blue,255},rotate=0](){};}}}

\begin{document}

\shorttitle{Low-$Re$ aerodynamics of finite-aspect-ratio swept wings} 
\shortauthor{K. Zhang \& K. Taira} 

\title{Aerodynamic characterization of low-aspect-ratio swept wings at $Re=400$} 

\author
 {
  Kai Zhang\aff{1}
  \corresp{\email{kai.zhang3@rutgers.edu}}
  \and
  Kunihiko Taira\aff{1}
  }

\affiliation
{
\aff{1}
Department of Mechanical and Aerospace Engineering, University of California, Los Angeles, CA 90095, USA
}

\maketitle

\begin{abstract}
\end{abstract}

\section{Introduction}
\label{sec:intro}

The low-Reynolds-number aerodynamics have been extensively studied over the past few decades due the interests in designing small-scale air vehicles and understanding biological flights.
At these scales, flows over wings exhibit complex flow physics comprised of unsteady separation, vortex formation, and wake interaction that are different from the high-Reynolds-number counterparts.
In fact, the aerodynamic characteristics of low-Reynolds-number flows exhibit strong nonlinearity arising from the rich vortex dynamics. 
\citet{liu2012numerical} and \citet{kurtulus2015unsteady} numerically investigated two-dimensional  unsteady flows over a NACA 0012 airfoil at a Reynolds number of 1000. Both studies highlighted the nonlinear behavior of the aerodynamic characteristics.
\citet{rossi2018multiple} assessed the Reynolds
number effects ($Re = 100 - 3000$) on the two-dimensional flows over a NACA 0010 airfoil and an ellipse at a fixed angle of attack of $30^{\circ}$. The presence of multiple bifurcations in the flow behavior and aerodynamic characteristics was reported.
More recently, \citet{menon2019aerodynamic} conducted a comprehensive study on the effects of two-dimensional airfoil
shapes and Reynolds number ($Re= 500- 2000$) on the aerodynamic characteristics. 
Their results showed sizeable and
rapid changes in the aerodynamic quantities with angle of attack, due to the presence of distinct flow phenomenon such as the Kármán vortex shedding and the formation of the leading-edge vortex.

Low-Reynolds-number flows around wings are further enriched by the end effects for finite-aspect-ratio wings. 
The nonlinear interactions among the tip vortices and the leading-/trailing-edge vortices lead to three-dimensional and aperiodic wakes \citep{winkelman1980flowfield,freymuth1987further,taira2009three,zhang2020formation}. 
Such vastly different flow physics from the analogous two-dimensional flows suggests that the understanding of fully three-dimensional analysis is necessary for practical wing designs at low Reynolds numbers. 
In our previous study, the wake dynamics of a NACA 0015 finite-aspect-ratio wings has been examined for a range of aspect ratios and angles of attack at $Re=400$ \citep{zhang2020formation}. The aerodynamic force coefficients of the finite-aspect-ratio wings were observed to be sigificantly lower than those of the two-dimensional airfoils even for wings with large aspect ratios. 
As we recently studied the flow over finite-aspect-ratio swept wings \citep{zhang2020laminar}, we observed that the sweep-induced midspan effects add another source of three dimensionality to the wake dynamics. 
However, the aerodynamic characteristics of the swept wings were not systematically reported in that study.

In this Technical Note, we present a database of aerodynamic force coefficients for finite-aspect-ratio swept  wings at a low Reynolds number of 400. 
The aerodynamic data are obtained from three-dimensional unsteady direct numerical simulations over a range of aspect ratios, angles of attack, and sweep angles. 
These data provides an improved understanding of the low-Reynolds-number aerodynamic characteristics of the canonical swept wings.

\section{Computational methods}

\begin{figure}
	\centering
	\includegraphics[width=0.75\textwidth]{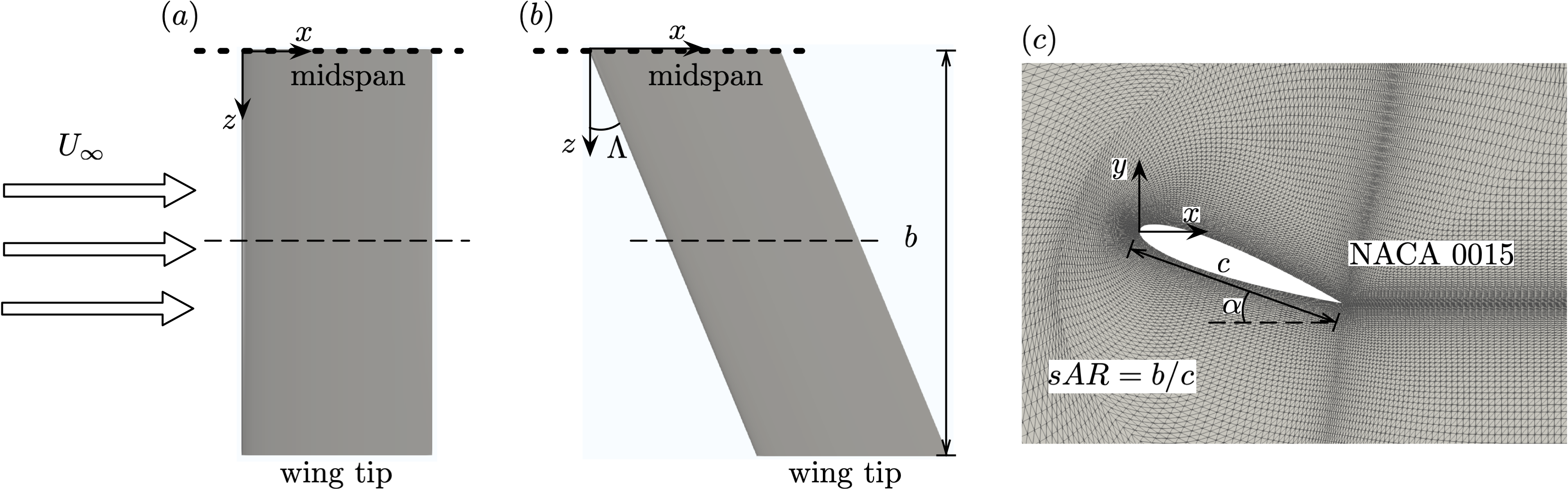}
	\caption{Case setup for $(a)$ unswept wing and $(b)$ swept wing. $(c)$ cross-sectional mesh at locations indicated by dashed line in $(a)$ and $(b)$.}
	\label{fig:scheme}
\end{figure}
For the present aerodynamic characterization, we simulate incompressible flows over finite-aspect-ratio swept wings with a NACA 0015 cross section. 
A schematic of the wing geometry is shown in figure \ref{fig:scheme}. 
The wings are subjected to uniform flow with velocity $U_{\infty}$ in the $x$ direction. The $z$ axis aligns with the spanwise direction of the unswept wing, and the $y$ axis points in the lift direction. 
For the swept cases, the wings are sheared towards the streamwise direction, and the sweep angle $\Lambda$ is defined as the angle between the $z$ axis and leading edge of the wing. We consider a range of sweep angles from $0^{\circ}$ to $45^{\circ}$.
The symmetry boundary condition is prescribed along the midspan (wing root). Denoting the half wing span as $b$, the semi aspect ratio is defined as $sAR=b/c$, where $c$ is the chord length, and is varied from 0.5 to 2. 
The Reynolds number, defined as $Re\equiv U_{\infty}c/\nu$ ($\nu$ is the kinematic viscosity of the fluid), is fixed at 400, at which the flow remains laminar.
The lift and drag coefficients are defined as $C_L=F_L/(\rho U^2bc/2)$ and $C_D=F_D/(\rho U^2bc/2)$, where $F_L$ and $F_D$ are the aerodynamic forces in $y$ and $x$ directions, respectively, and $\rho$ is the fluid density.

The incompressible solver \emph{Cliff} (in \emph{CharLES} software package, Cascade Technologies, Inc.) is used for simulating the flows over wings using direct numerical simulations. This solver employs a collocated, node-based finite-volume method to simulate the flows with second-order spatial and temporal accuracies \citep{ham2004energy,ham2006accurate}. The computational domain and mesh set-ups in this study follow our previous works \citep{zhang2020formation,zhang2020laminar}, which have been extensively validated.

\section{Results}
\subsection{Wake dynamics}
\label{sec:wake}
\begin{figure}
	\centering
	\includegraphics[width=0.9\textwidth]{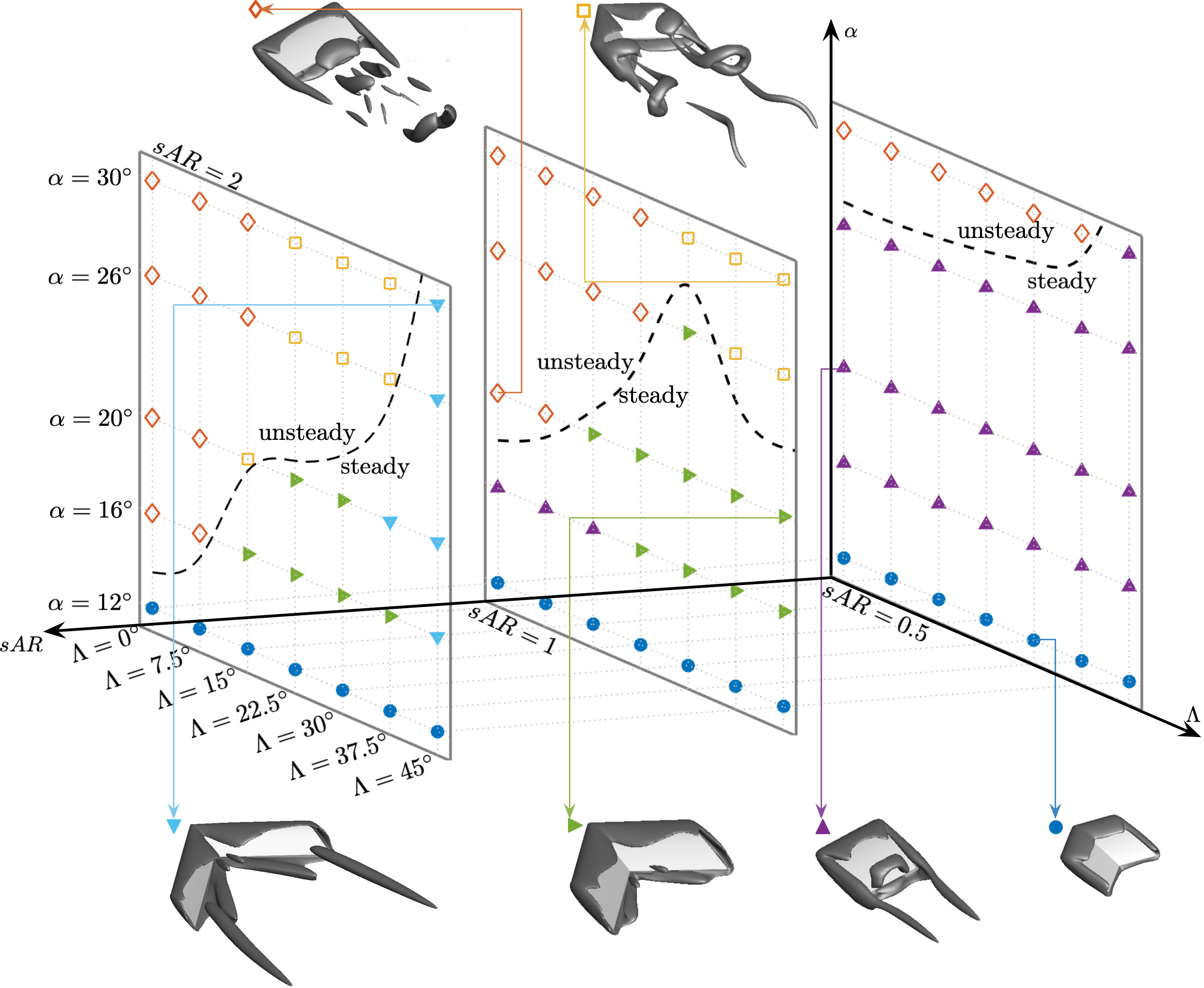}
	\caption{Classification of flows around finite-aspect-ratio wings. \protect\markerone: steady flows; \protect\markertwo: steady flow due to tip effects; $\MyDiamond[draw={rgb,255:red,217; green,83; blue,25},line width=0.3mm, fill=white]$: unsteady shedding near midspan; \protect\markerthree: steady flow due to midspan effects; \protect\markerfive: unsteady shedding near wing tip; \protect\markerfour: steady flow with streamwise vortices. The dashed lines denote the approximate boundaries between steady (filled symbols) and unsteady (empty symbols) flows. The vortical structures are visualized by isosurfaces of $Qc^2/U_{\infty}^2=1$ for representative cases. }
	\label{fig:regime}
\end{figure}

We begin the discussions by presenting an overview of the wake dynamics of the swept finite-aspect-ratio wings.
A classification of the wakes is presented in figure \ref{fig:regime}, with representative vortical structures shown for selected cases.
For wings with $\alpha\lesssim 12^{\circ}$, the wake over a NACA 0015 airfoil at $Re=400$ remains stable. Steady flows without significant formation of tip vortices (\protect\markerone), are observed regardless of the aspect ratio and sweep angle.
For higher angles of attack, the wake dynamics are influenced by the complex interplay between the tip effects and the midspan effects. 
The tip effects are responsible for the formation of steady wakes with low aspect ratios and low sweep angles (\protect\markertwo). In these cases, the downwash induced by the tip vortices suppresses the roll-up of the vortex sheet on the suction side of the wing  \citep{taira2009three,devoria2017mechanism}.
With an increase in aspect ratio, the effects of the tip vortices become relatively weaker away from the tip. This allows for the roll-up of the leading-edge vortex sheets, resulting in unsteady vortex shedding near the midspan ($\MyDiamond[draw={rgb,255:red,217; green,83; blue,25},line width=0.3mm, fill=white]$). Compared with $sAR=0.5$, the stability boundaries of the wake for $sAR=1$ and 2 shift toward lower angles of attack for low-sweep wings.

For wings with larger aspect ratios and larger sweep angles, the tip vortices are weaker than those for lower sweep wings, and the midspan effects become profound in shaping the wakes dynamics.
The midspan effects are associated with the formation of a pair of vortical structures on the suction side of the midspan. 
These vortical structures are aligned at an angle of $180^{\circ}-2\Lambda$. For $\Lambda\neq 0^{\circ}$, each of the vortical structures is subjected to the downward velocity induced by its symmetric peer on the other side of the midspan. Such mechanism stabilizes the wake over a considerable number of cases (\protect\markerthree).
The formation of the vortical structures near the midspan is also beneficial to the aerodynamic performance, as it will be discussed in detail in the following sections.

The downward velocity described above is strong near the midspan and weak towards the outboard sections of the wing.
For swept wings with large aspect ratios, unsteady vortex shedding develops locally near the tip region, while the midspan region still remains steady. The resulting flows resemble the ``tip stall" phenomenon \citep{black1956flow,visbal2019effect}, and prevail for swept wings of $sAR=1-2$ with high angles of attack (\protect\markerfive). 
For wings of $sAR=2$ with high sweep angles ($\Lambda=37.5^{\circ}-45^{\circ}$), the unsteady tip shedding further transitions to another type of steady flow, with the formation of the streamwise vortices (\protect\markerfour).
We refer the readers to our previous study \citep{zhang2020laminar} for a thorough discussion on the wake dynamics of swept wings.

\subsection{Lift coefficients}
\label{sec:lift}

\begin{figure}
	\includegraphics[width=1\textwidth]{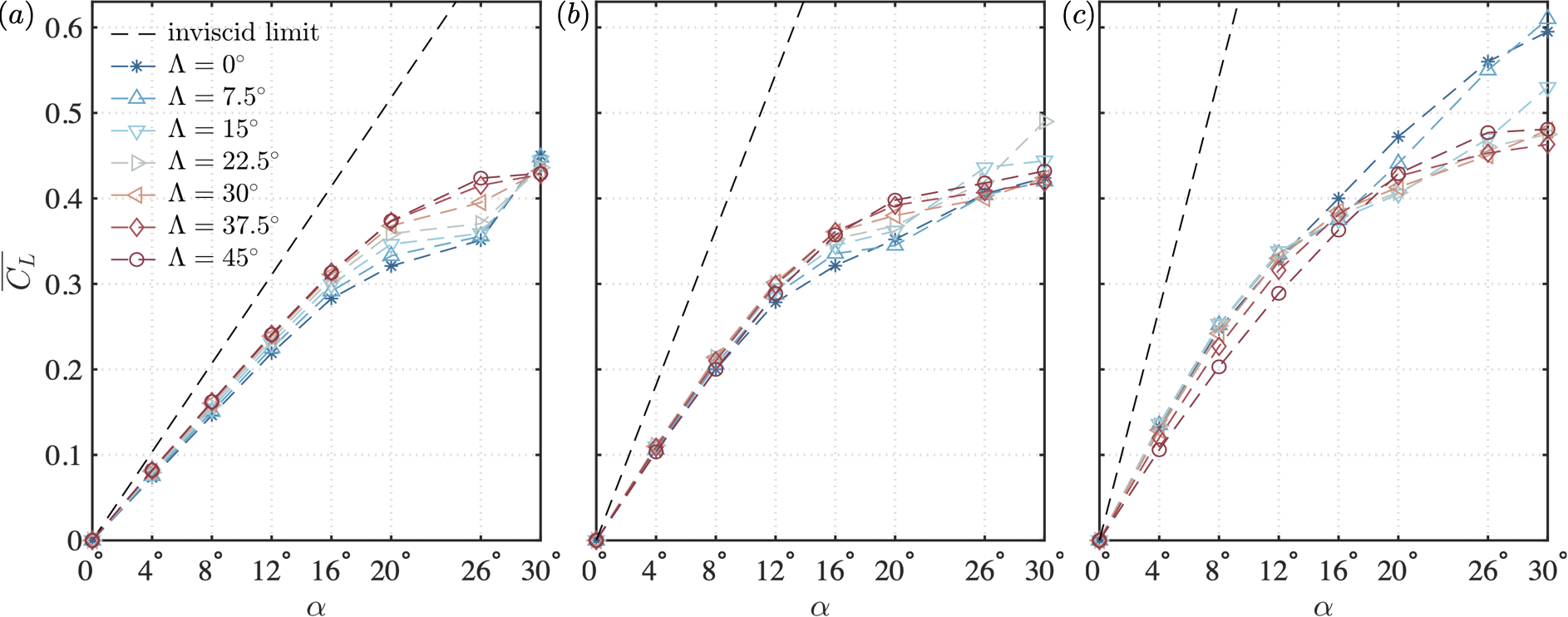}
	\caption{Time-averaged lift coefficients for $(a)$ $sAR=0.5$, $(b)$ $sAR=1$ and $(c)$ $sAR=2$.}
	\label{fig:lift}
\end{figure}
The lift force coefficients of the wings with $sAR=0.5$, 1 and 2 are presented in figure \ref{fig:lift}. Also plotted are the inviscid limit for the lift of low-aspect-ratio unswept wings in incompressible flow \citep{helmbold1942unverwundene}:
\begin{equation}
C_L = \displaystyle{\frac{2\pi \alpha}{\sqrt{1+(1/sAR)^2}+1/sAR}}.
\end{equation}

The lift coefficients of the finite-aspect-ratios wings are significantly smaller than the inviscid limit for all three aspect ratios.
Compared to a similar characterization \citep{taira2009three} for flat-plate wings at $Re=300$, we find that the airfoil shape is influential even at these low Reynolds numbers.
For $sAR=0.5$, the sweep has a positive effect on the lift coefficients for $\alpha \lesssim 26^{\circ}$. 
The vortical lift enhancement with the sweep angle is due to the vortical structures near the midspan, as discussed in section \ref{sec:wake}.
As the flow transitions to unsteady shedding at $\alpha=30^{\circ}$, the lift coefficients undergo an abrupt jump for wings with low sweep. For these unsteady flows, the lift coefficients decrease slightly with the sweep angle.

As the aspect ratio increases to $sAR=1$, for $\alpha\lesssim 12^{\circ}$, the lift coefficients across different sweep angles remain close to each other, and increase almost linearly with the angle of attack with a steeper slope than that of the analogous cases with $sAR=0.5$.
For $\alpha\approx 16^{\circ}-20^{\circ}$, the favorable effect of sweep on the lift coefficients becomes more noticeable. The increase of $\overline{C_L}$ with $\Lambda$ saturates at high sweep angles.
For higher angles of attack ($\alpha\approx 26^{\circ}-30^{\circ}$), the lift coefficients no longer exhibit a monotonic relationship with the sweep angle. 
Instead, high $\overline{C_L}$ is observed for $\Lambda=15^{\circ}$ at $\alpha=26^{\circ}$, and $\Lambda=22.5^{\circ}$ at $\alpha=30^{\circ}$.

For $sAR=2$, the lift coefficients decrease with increasing sweep angle for wings with low angles of attack ($\alpha\lesssim 16^{\circ}$). 
The adverse effect of sweep on lift at $sAR=2$ (contrary to the positive effect at $sAR=0.5$) is due the fact that the additional generation of vortical lift is limited to the midspan region, while the elongated outboard region is featured by lower sectional lift. 
However, at higher angles of attack, the lift coefficients for $\Lambda=45^{\circ}$ surpass those of moderate sweep angles ($\Lambda=15^{\circ}-37.5^{\circ}$), although they are significantly smaller than those for $\Lambda=0^{\circ}-7.5^{\circ}$. 
Compared with $sAR=0.5$ and 1, the lift coefficients for $sAR=2$ wings are generally higher. An exception of the this trend is observed for $\Lambda=45^{\circ}$ wings for $\alpha=0^{\circ}-16^{\circ}$, where the lift coefficients remain almost the same with those of the $sAR=1$ wings.

\subsection{Drag coefficients}
\begin{figure}
	\includegraphics[width=1\textwidth]{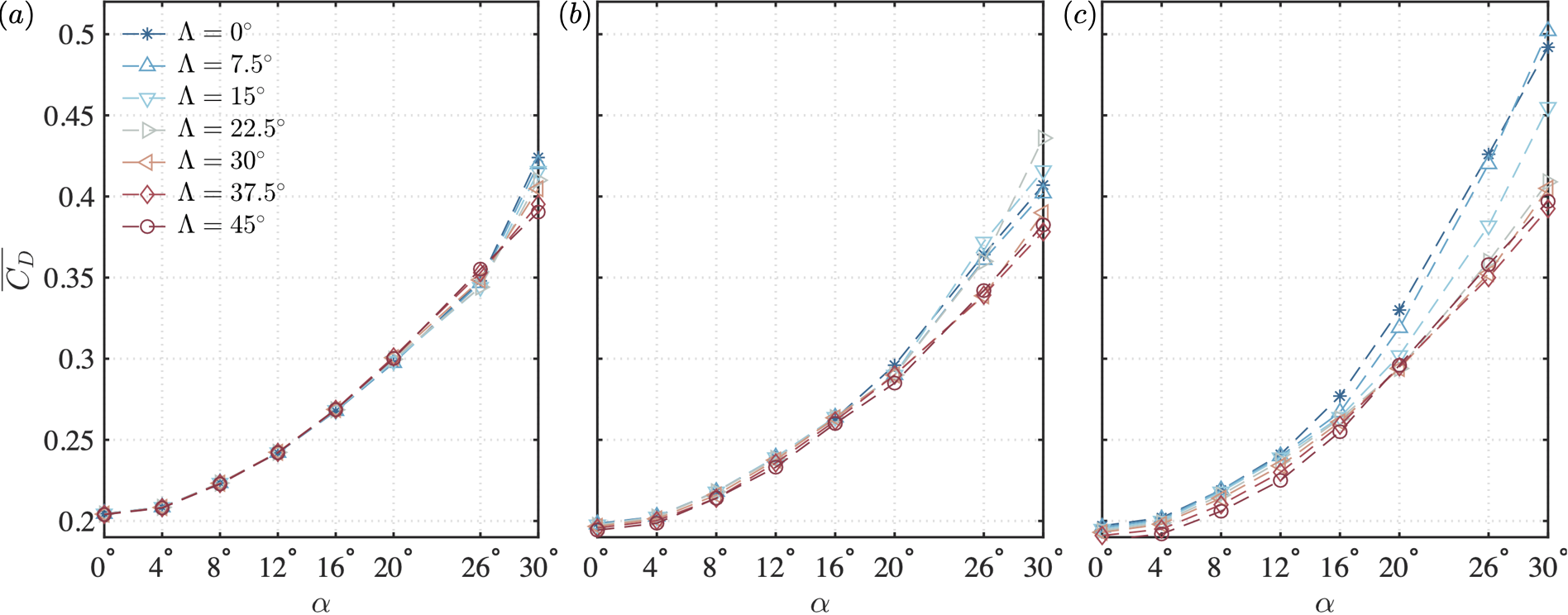}
	\caption{Time-averaged drag coefficients for $(a)$ $sAR=0.5$, $(b)$ $sAR=1$ and $(c)$ $sAR=2$.}
	\label{fig:drag}
\end{figure}
The drag coefficients of the wings exhibit an quadratic growth with angle of attack over the studied range, as shown in figure \ref{fig:drag}. 
For the low aspect ratio of $sAR=0.5$, the difference in drag coefficients among cases with different sweep angles is not noticeable until $\alpha=20^{\circ}$. 
At $\alpha=26^{\circ}$, $\overline{C_D}$ increases with the sweep angle.
As the flow destabilizes at $\alpha=30^{\circ}$, similar to the lift coefficients shown in figure \ref{fig:lift}($a$), the drag coefficient becomes negatively affected by the sweep angle.
Compared to the drag coefficients at $sAR=0.5$, those at $sAR=1$ are generally smaller for $\alpha\lesssim 20^{\circ}$.
For these cases, the drag decreases with increasing $\Lambda$, although the difference among different sweep angles remains small. 
At higher angles of attack ($\alpha=26^{\circ}-30^{\circ}$), the drag coefficients of wings with $\Lambda=0^{\circ}-22.5^{\circ}$ are significantly higher than those with $\Lambda=30^{\circ}-45^{\circ}$. 
Similar to $sAR=1$, the drag coefficients at $sAR=2$ also decreases with increasing sweep angle. 
However, the difference in $\overline{C_D}$ among different sweep angles becomes larger even at low angles of attack. 
At higher angles of attack, the drag coefficients of wings with low sweep angles grow much faster with $\alpha$  than those with high sweep angles.

\subsection{Lift-to-drag ratios}

\begin{figure}
	\includegraphics[width=1\textwidth]{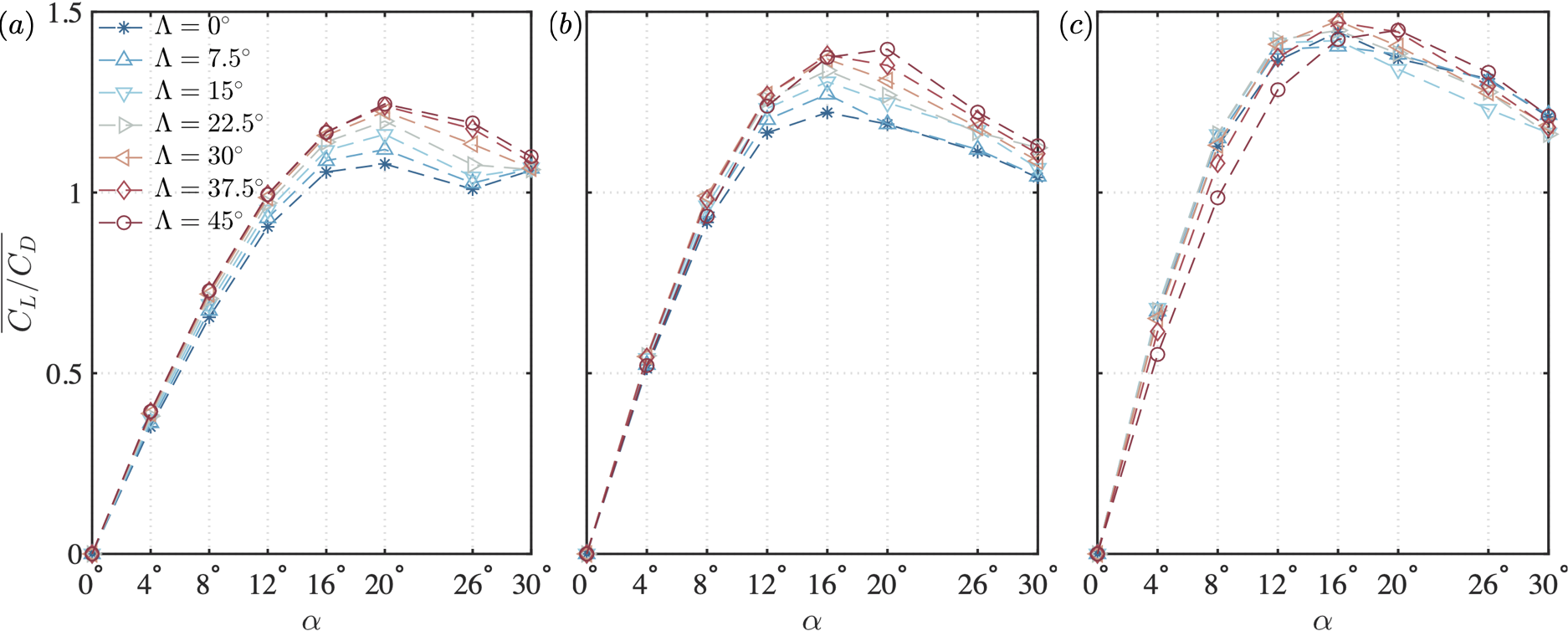}
	\caption{Time-averaged lift-to-drag ratio for $(a)$ $sAR=0.5$, $(b)$ $sAR=1$ and $(c)$ $sAR=2$.}
	\label{fig:ratio}
\end{figure}

The time-averaged lift-to-drag ratios  are compiled in figure \ref{fig:ratio}. 
The $\overline{C_L/C_D}$ generally improves with increasing aspect ratio.
However, due to the low-Reynolds-number nature of the flow, the lift-to-drag ratio remains below 1.5 for the cases considered herein.
At $sAR=0.5$, the lift-to-drag ratio increases with the angle of attack up to $\alpha\approx 20^{\circ}$, at which the maximum $\overline{C_L/C_D}$ is achieved. 
The lift-to-drag ratio increases with the sweep angle, due to the lift enhancement mechanism of the midspan effects discussed in \S \ref{sec:wake}.
The positive effect of sweep angle on  $\overline{C_L/C_D}$ is also observed for wings with $sAR=1$. The maximum $\overline{C_L/C_D}$ for $sAR=1$ is achieved at $\alpha\approx 16^{\circ}$ for $\Lambda=0^{\circ}-37.5^{\circ}$, and at $\alpha\approx 20^{\circ}$ for $\Lambda=45^{\circ}$. 
For $sAR=2$ at low angles of attack ($\alpha\lesssim 12^{\circ}$), the lift-to-drag ratios of the $\Lambda=45^{\circ}$ wings are significantly lower than those with lower sweep angles. At higher angles of attack ($\alpha=20^{\circ}-30^{\circ}$), $\overline{C_L/C_D}$ of the $\Lambda=45^{\circ}$ wing are only slightly higher than the rest of the cases. 
This suggests that care should be taken in selecting the right type of wings if midspan lift enhancement is to be taken advantage of with finite-aspect-ratio swept wings.

\section{Conclusions}
\label{conclusions}
We have performed unsteady three-dimensional direct numerical simulations to study the aerodynamic characteristics of finite-aspect-ratio swept wings with a NACA 0015 cross-section at a chord-based Reynolds number of 400. 
The effects of the sweep angle ($\Lambda=0^{\circ}-45^{\circ}$) on the aerodynamic force coefficients were examined for finite-aspect-ratio wings ($sAR=0.5$, 1, and 2) over a wide range of angles of attack ($\alpha=0^{\circ}-30^{\circ}$).
The unsteady laminar separated flows exhibit complex aerodynamic characteristics with respect to these parameters.
The introduction of sweep enhances lift for wings with low aspect ratios of $sAR=0.5$ and 1, due to the  lift generated by the vortical structures near the midspan. 
For these cases, the dependence of drag coefficients on the sweep angle is less noticeable, particularly at lower angles of attack. 
The lift-to-drag ratios for low-aspect-ratio wings increase with the sweep angle.
However, such favorable effects of sweep angle on the lift coefficients and lift-to-drag ratios are not observed for wings with higher aspect ratios, where the midspan effects are relatively weaker over the wing span. 
The results herein provide a laminar aerodynamic characterization of the low-aspect-ratio wings and complement the unsteady low-Reynolds-number aerodynamic database with highlight on the effect of sweep.

\section*{Acknowledgments}
We acknowledge the US Air Force Office of Scientific Research (Program Managers: Dr.~Gregg Abate and Dr.~Douglas Smith, Grant number: FA9550-17-1-0222) for funding this project. We thank Ms.~Shelby Hayostek, Prof.~Michael Amitay, Dr.~Wei He, Mr.~Anton Burtsev and Prof.~Vassilios Theofilis for insightful discussions.

\bibliography{sample}
\bibliographystyle{jfm}

\end{document}